\documentclass[12pt]{iopart}
  \expandafter\let\csname equation*\endcsname\relax
  \expandafter\let\csname endequation*\endcsname\relax
\usepackage{amsmath,amssymb}

\usepackage{graphicx}

\newcommand{\be}{\begin{equation}}
\newcommand{\ee}{\end{equation}}
\newcommand{\bea}{\begin{eqnarray}}
\newcommand{\eea}{\end{eqnarray}}
\newcommand{\ba}{\begin{array}}
\newcommand{\ea}{\end{array}}

\begin{document}

\title{Chaotic properties of Coulomb-interacting circular billiards}

\author{J. Solanp\"a\"a$^{1,2}$, J. Nokelainen$^2$, 
P. Luukko$^2$ and E. R{\"a}s{\"a}nen$^{1,2}$}
\address{$^1$ Department of Physics, Tampere University of Technology,
FI-33101 Tampere, Finland}
\address{$^2$ Nanoscience Center, Department of Physics, University of Jyv{\"a}skyl{\"a}, FI-40014 Jyv{\"a}skyl{\"a}, Finland}

\begin{abstract}
We apply a molecular dynamics scheme to analyze classically
chaotic properties of a two-dimensional circular billiard system
containing two Coulomb-interacting electrons. As such, the system
resembles a prototype model for a semiconductor quantum dot.
The interaction strength is varied from the noninteracting limit with zero potential energy
up to the strongly interacting regime where the relative kinetic energy
approaches zero. At weak interactions the bouncing maps show jumps between
quasi-regular orbits. 
In the strong-interaction limit we find an analytic expression
for the bouncing map. Its validity in the general case is assessed by
comparison with our numerical data. To obtain a more quantitative view
on the dynamics as the interaction strength is varied, we compute and analyze
the escape rates of the system. Apart from very weak or strong interactions,
the escape rates show consistently exponential behavior, thus suggesting
strongly chaotic dynamics and a phase space without significant 
sticky regions within the considered time scales.
\end{abstract}


\maketitle

\section{Introduction}

Classical billiard systems have attracted continuous
interest for several decades due to their applicability to
demonstrate chaotic dynamics through (semi)analytic and numerical
calculations~\cite{gutzwiller,stockmann,nakamura}.
On the other hand, laboratory experiments on, e.g., microwave
billiards~\cite{stockmann}, quantum dots~\cite{nakamura},
and more recently even graphene~\cite{graphene} have
rapidly extended the interest in chaos across different fields in
physics. Along this development, billiard systems have become
a key element in the studies of classical and
quantum chaos both theoretically and experimentally.

Most billiard studies have focused on {\em single-particle}
properties of systems ranging from regular (integrable)
to chaotic (nonintegrable) systems, including also
pseudointegrable billiards~\cite{pseudo} such as regular billiards
with singular scatterers inside the system. Two-particle
billiards have been studied with hard-sphere contact
interactions in, e.g., rectangular~\cite{awazu} and
mushroom-shaped~\cite{lansel} cavities. Also two-particle billiards with Yukawa
interactions have been studied in one-dimensional systems (1D)~\cite{yukawa_1d,yukawa_coulomb_soft}
and two-dimensional systems such as circular billiards in
both classical~\cite{yukawa_circle} and
quantum~\cite{yukawa_circle_quantum} cases. To the best of our
knowledge, such studies with Coulomb interactions -- and
with the focus on classical chaotic properties -- have been
restricted to two-dimensional (2D) harmonic
oscillators~\cite{rost,aaberg,drouvelis1,drouvelis2} including
an anharmonic oscillator~\cite{sebastian}. Exceptions to this
class are periodic systems~\cite{knauf_periodic} as well
as rectangular billiards in magetic fields~\cite{aichinger}
studied with molecular dynamics (MD).

The MD scheme is a computationally
efficient approach to many-particle billiards that, in principle,
can be extended to large systems without compromising the
numerical complexity of the long-range Coulomb interaction.
It is noteworthy that the Coulomb interaction is a
physically meaningful choice when considering similar systems
in, e.g., quantum-dot physics~\cite{nakamura,kouwenhoven,reimann}.
Experimentally, vertical or lateral semiconductor quantum dots can be
tailored at will with respect to the system shape, size, and number of
confined electrons. In this respect examination on the interaction
effects in few-electron billiards have immediate relevance to physical
applications.

Here we adopt the MD approach to analyze the classical chaoticity
of a 2D circular hard-wall billiards with two Coulomb-interacting
electrons. This particular system is chosen under examination as it
represents, alongside a harmonic oscillator, the simplest prototype
model for a quantum dot. Secondly, the {\em noninteracting}
properties of the system are well known~\cite{gutzwiller}. We may
also expect to find analytic, approximate expressions for
the bouncing map in the {\em strong-interaction limit}.
In the intermediate
regime, the system is expected to exhibit chaotic behavior.
Due to these features the system provides a well-grounded path into
examinations of both classical and
quantum chaos in Coulomb-interacting billiard systems. We point out that
soft billiards are better known in this respect; for example, the
two-electron circular harmonic oscillator is regular and becomes mixed
(partly regular, partly chaotic) if
ellipticity is added in the external potential~\cite{drouvelis2}.

We can always introduce an open billiards corresponding to a given closed billiards by generating
holes in the boundary via which the particle(s) can escape the table. The escape probability at some infinitesimal time interval 
(or at a certain number of collisions) is called the {\em escape rate}.
The form of the escape-rate distribution is governed by the structure of the phase
space~\cite{escape_phase_space_dependence} 
and the position(s) of the hole(s)~\cite{effect_of_positions_of_holes}. If the phase space has
sticky regions, i.e., regions where a (possibly chaotic) trajectory gets stuck for a
long period of time, the escape rate and survival probability turn out to have an algebraic tail
as time tends to infinity~\cite{kam_sticky2,sticky_asymptotics,sticky_asymptotics2,decay_meiss,decay_meiss2,decay_hanson,decay_types}.
On the other hand, if the phase space is fully chaotic and non-sticky, the escape rate is
asymptotically exponential. Sticky regions can result from several origins.
For example, internal stickiness -- not due to 
Kolmogorov--Arnold--Moser (KAM) tori -- can be induced
by marginally stable periodic orbits~\cite{decay_types_stadium,mupo_sticky}. External stickiness, on the other hand,
is caused by sticky KAM tori~\cite{kam_sticky1,kam_sticky2,altmann}, albeit not all KAM tori are sticky~\cite{nonsticky_kam}.
Different types of stickiness have been recently reviewed by
Bunimovich and Vela-Arevalo in Ref.~\cite{many_faces}.

The paper is organized as follows. In Sec.~\ref{methods} we briefly
introduce the system and our time-propagation scheme.
In Sec.~\ref{bouncing_maps} we show bouncing maps that demonstrate
clear signals of chaotic behavior through a large range of the interaction
strength. At weak interactions, bouncing maps are found to jump between
quasi-regular trajectories. In Sec.~\ref{strong} we analyze in detail
the strong-interacting limit and find an approximate expression for the bouncing map.
The expression agrees with the numerical results, and at weaker interactions it becomes
only approximate. Finally, in Sec.~\ref{escape} we assess the degree of chaoticity by
considering escape rates out of the system. Apart from very weak interactions,
we find exponential escape in a wide range of the interaction strength.
This indicates strongly uncorrelated trajectories and thus chaotic behavior.
The paper is summarized in Sec.~\ref{summary}.

\section{System and methodology}\label{methods}

We consider two Coulomb-interacting electrons
in a circular hard-wall potential. The collisions with the
boundary are elastic and the system is described by the Hamiltonian
\be
\label{eqn:hamiltonian}
H=\frac{1}{2} \left(v_1^2+v_2^2\right)+\frac{\alpha}{\left|\mathbf{r}_{1}-\mathbf{r}_{2}\right|}
\ee
in Hartree atomic units (a.u.) ($\hbar=e=m_e=(4\pi\epsilon_0)^{-1}=1$).
Here $\mathbf{r}_i$ is the position vector of the $i$th electron from
the center of the system, and $\alpha$ is a parameter that
determines the interaction strength.
In all our simulations the total energy of the system is fixed to
$E=1$ and the radius of the circle to $R=1/2$.
The interaction strength is restricted to
$0 \leq \alpha \leq 1$, where $\alpha=0$ corresponds to noninteracting
electrons, and $\alpha=1$ corresponds to electrons being localized at
the opposite sides of the circle with zero motion.

To propagate the electrons we use molecular dynamics
with the velocity Verlet \cite{velocityverlet} algorithm which is as a symplectic and time-reversible algorithm
suitable for the study of (possibly chaotic) Hamiltonian systems.
A higher order integrator is not necessary for the system under 
consideration: the numerical uncertainty resulting from a finite time step is dominated
by collisions with the boundary instead of the integration of Hamilton's equations of motion.
In the velocity Verlet algorithm the positions
and velocities of each electron are calculated from
\bea
\mathbf{r}(t+\Delta t) & = & \mathbf{r}(t)+\mathbf{v}(t)  \Delta t+\frac{1}{2}\mathbf{a}(t) \Delta t^2;\label{vv1}\\
\mathbf{v}(t+\Delta t/2) & = & \mathbf{v}(t)+\frac{1}{2}\mathbf{a}(t) \Delta t;\label{vv2}\\
\mathbf{a}(t+\Delta t) & = & \sum\limits_i\mathbf{F}_{i}\left[\mathbf{r}(t+\Delta t)\right];\label{vv3}\\
\mathbf{v}(t+\Delta t) & = & \mathbf{v}(t+\Delta t/2)+\frac{1}{2}\mathbf{a}(t+\Delta t)  \Delta t.\label{vv4}
\eea
We define $\cos\theta$ and $s$ as the generalized coordinates
describing the collisions with the boundary. $\theta$ is
the angle between the velocity vector of the incoming electron and
the tangent of the boundary, so that
$\theta<\pi/2$ and $\theta>\pi/2$ correspond to counterclockwise
and clockwise traveling directions, respectively.
Here $s\in]-\pi/2,\,\pi/2]$ is the oriented arc length from the chosen origin.

\section{Results}

\subsection{Bouncing maps}
\label{bouncing_maps}

\begin{figure}
\centering
\includegraphics[width=0.7\columnwidth]{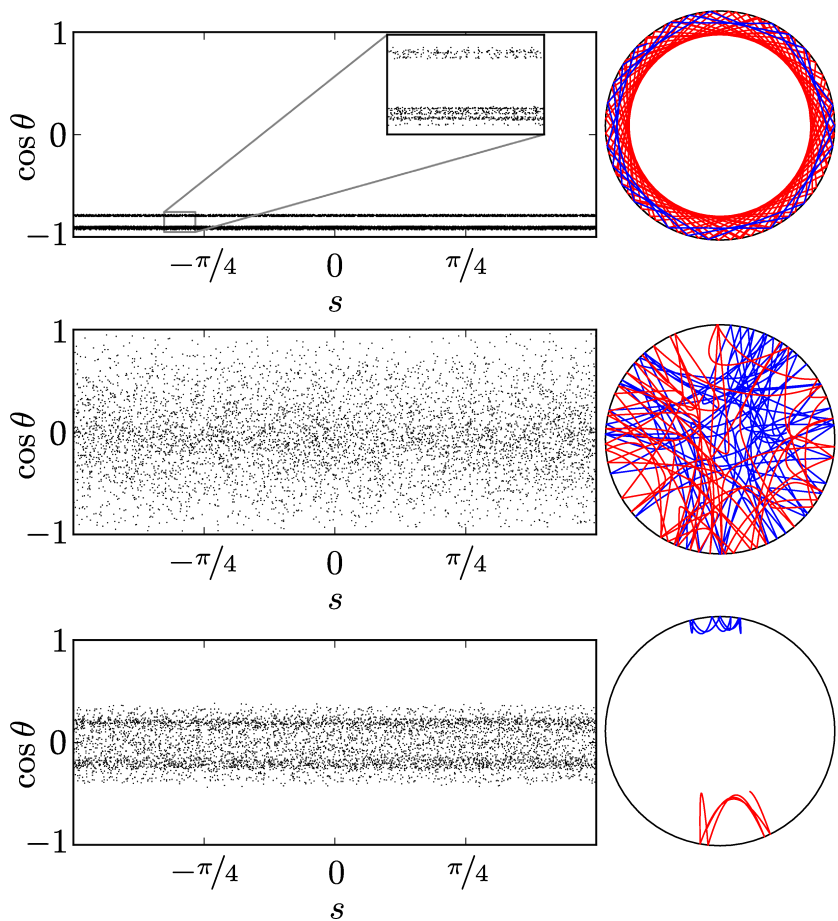}
\caption{Examples of bouncing maps (of one of the electrons) and trajectories for a two-electron
circular billiards with different interaction strengths:
$\alpha=10^{-5}$ (up), $\alpha=0.2$ (middle), $\alpha=0.7$ (bottom). Only a small section of the
trajectories corresponding to the bouncing maps are shown.}
\label{fig1}
\end{figure}

In Fig.~\ref{fig1} we show examples of bouncing maps and electron
trajectories for a two-electron circular billiard with
different interaction strengths $\alpha$. In this case the
bouncing maps consist of 14 000 ($\alpha=10^{-5}$)
and 5500 ($\alpha=0.2$ and $0.7$) collisions with the boundary.
The noninteracting circular system with $\alpha=0$ is a well-known
example of regular billiards~\cite{gutzwiller} represented by straight
lines in the map (constant bouncing angle) and straight trajectories forming
a star-shaped pattern, where the inner part of the circle remains empty.
When $\alpha=10^{-5}$ we find emerging deviations from this limit as visualized
in the inset of the upper panel of Fig.~\ref{fig1}. When the electrons
pass each other the interaction is pronounced and we may find
``jumps'' from one quasi-regular trajectory to another one
(close-lying parallel lines in the inset).

In the intermediate-interaction range (middle panel of Fig.~\ref{fig1})
the chaoticity of the system is clear, so that the
bouncing map rapidly becomes completely filled. As expected, the
distribution of the bouncing map is centered at $\theta=\pi/2$,
so that, {\em on the average}, the electrons hit the boundary along the
normal vector.

If $\alpha$ is increased above $\alpha\sim 0.5$ we find that for some trajectories the
maximum of the probability distribution for $\theta$ splits into two.
This is visible in the bottom panel in Fig.~\ref{fig1} for $\alpha=0.7$.
However, the splitting is smoothed out when a large ensemble of
trajectories is taken into account. When
$\alpha$ is increased further, the system gradually becomes
(quasi)regular and eventually the bouncing map reduces into a
one-dimensional curve.
In the following we carry out analytic calculations in the strong-interaction
limit.

\subsection{Strong-interaction limit}\label{strong}

\begin{figure}
\centering
\includegraphics[width=0.5\columnwidth]{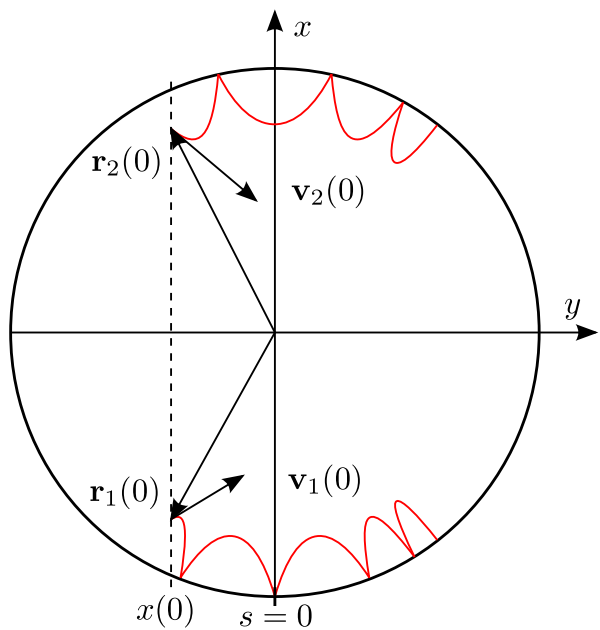}
\caption{Two electrons oscillating at the opposite sides
of the circle when the interaction is strong ($\alpha\gtrsim 0.9$).
The initial positions and velocities are shown.
}
\label{fig3}
\end{figure}

In the strong-interaction limit $\alpha\rightarrow 1$,
the two-electron dynamics shows regular characteristics.
The electrons are confined at opposite sides of the circle as
visualized in Fig.~\ref{fig3}. Here
we focus on the special case with total angular momentum
$L=0$ which is conserved due to the rotational symmetry.
Hence, according to the choice of coordinate axes in
Fig.~\ref{fig3} we may approximate $x \equiv r_{1,x} \approx r_{2,x}$.
We point out, however, that the $y$ coordinate does not usually
have mirror-symmetry. Note also the position
of $s=0$ at $x=0$ in Fig.~\ref{fig3}, so that $s\in[-s_\text{max},s_\text{max}]$.


As the second approximation,
the electron velocity perpendicular to the edge when a collision takes place,
$v_\perp$, can be taken as a constant, i.e., it is approximately the
same for all possible values of $s$ at all times.
After a straightforward geometrical
analysis, taking into account the conservation of $E$ and $L$, we
can calculate the tangential velocity $v_{\shortparallel}(s)$ during the collision
and further the cosine of the bouncing angle from 
$\tan \theta = \) \(v_{\perp}\)/\(v_{\parallel}$.
Thus, we obtain the following strong-interaction approximation for
the bouncing map of electron 1:
\be
\cos\theta_1(s)
= \pm\left(1 +
\frac{2K + 2U_0 - 2U_\mathrm{a} - L^2/R^2}
{U_\mathrm{b} - U_\mathrm{c}(s) + L^2/R^2}
\right)^{-1/2},
\label{eqn:theta}
\ee
where
$K = \left[v_{1,x}^2(0) + v_{1,y}^2(0)\right]/2$ is the initial kinetic
energy of the electron, $L=x(0) v_{1,y}(0) - y_1(0) v_{1,x}(0)$ is
its initial angular momentum (note that $x=x_1\approx x_2$ according to our
approximation above), $U_0 = \alpha/[|y_1(0)| + |y_2(0)|]$
is the initial potential energy, and $R=1/2$ is the radius of the
circle. Furthermore,
Eq.~(\ref{eqn:theta}) has three potential energy components
that have expressions
\bea
U_\textrm{a} &=& \frac{\alpha}{|y_2(0)| + \sqrt{R^2 - x(0)^2}},
\label{Ua}\\
U_\text{b} &=& \frac{\alpha}{2\sqrt{R^2-x(0)^2}},
\label{Ub}\\
U_\text{c}(s) &=& \frac{\alpha}{2R\cos{(s/R)}}.
\label{Uc}
\eea
They correspond to the following situations where both
$E$ and $L$ are conserved and $x_1 = x_2$.
Firstly, $U_\textrm{a}$ is the potential energy corresponding to a situation
where electron 1 touches the boundary at $t=0$ and electron 2 has its
initial position and velocity.
Secondly, $U_\textrm{b}$ corresponds to a situation
where both electrons touch the boundary at $t=0$,
and $v_1$ is the same as in the previous (first) case.
Finally, $U_\textrm{c}$ corresponds to a situation
where both electrons touch the boundary at $s$ at unknown time,
having the same $v_\perp$ as in the previous (second) case.

\begin{figure}
\centering
\includegraphics[width=0.7\columnwidth]{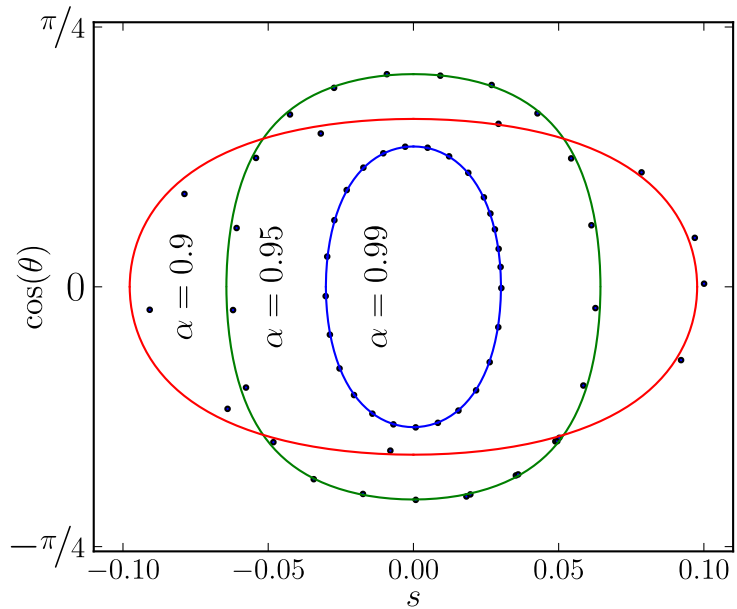}
\caption{Analytic result for the large-$\alpha$ limit
[Eq.~(\ref{eqn:theta})] of the bouncing map (solid lines)
for $\alpha=0.99$, $0.95$, and $0.9$. The corresponding
simulated, i.e., numerically exact values are shown
by points.
}
\label{fig4}
\end{figure}

Figure~\ref{fig4} shows the results from Eq.~(\ref{eqn:theta})
for $\alpha=0.99$, $0.95$, and $0.9$ (solid lines).
The {\em simulated}, i.e., the numerically exact values, are
shown by points for comparison. As $\alpha$ is decreased we find gradual
deviation from the simulated data. At $\alpha=0.9$
the deviation is already clearly visible. We may thus
state that Eq.~(\ref{eqn:theta}) provides a reasonable
approximation for the bouncing map at $\alpha\gtrsim 0.9$.
This threshold slightly depends on the initial conditions;
the examples in Fig.~\ref{fig4} are chosen such that
the deviations between the analytic expression and the
numerical data are large.


\subsection{Escape rates}\label{escape}

Next we examine how the dynamics of the system changes as we
move from the noninteracting ($\alpha=0$) to the strongly
interacting ($\alpha\to 1$) limit. A full description of the seven-dimensional
phase space, for example by means of Poincar\'e section, would be difficult.
Therefore, we consider escape rates of the system by placing holes in the boundary.
As already discussed in the introduction, systems with sticky regions in phase space
have power-law asymptotics of the escape rate distribution, whereas in fully
chaotic systems without stickiness the escape-rate distribution
is exponential as $t\rightarrow\infty$.
We remind that escape rates are commonly governed by stickiness 
rather than regular/chaotic components of the phase space.
Therefore, we cannot make a complete assessment of the structure of 
the phase space, especially not close to the limits 
$\alpha=0$ and $\alpha\to 1$.

We set 10 holes in the boundary covering together $1/50$ of the
boundary length -- the same fraction as in Ref.~\cite{bauer_bertsch}.
The escape rates are considered as a function of the total number of
collisions $n$, i.e., the sum of collisions of both particles, rather than the propagation time, since the
characteristic time scale strongly depends on the interaction strength $\alpha$.
For each $\alpha$ we compute $2.5\ldots 6\times 10^5$ respective trajectories
with random initial conditions and store the number of boundary
collisions before the escape. Initial conditions are randomized in the following way:
First we pick random initial
conditions for the particles in the energetically allowed ($E_\text{total}=1$) part of the configuration space
and then we distribute the remaining energy evenly as kinetic energy among particles. Also the directions of the velocities are randomized.
The escape rate $P(n)$ is defined as the ratio of the
number of trajectories escaping at $n$th collision to the number of trajectories in the initial ensemble.
The time steps are chosen in the range $\Delta t=10^{-8} \ldots 10^{-7}$,
so that the convergence is ensured in every calculation,
while the numerical efficiency is maximized.

\begin{figure}
\centering
\includegraphics[width=0.7\columnwidth]{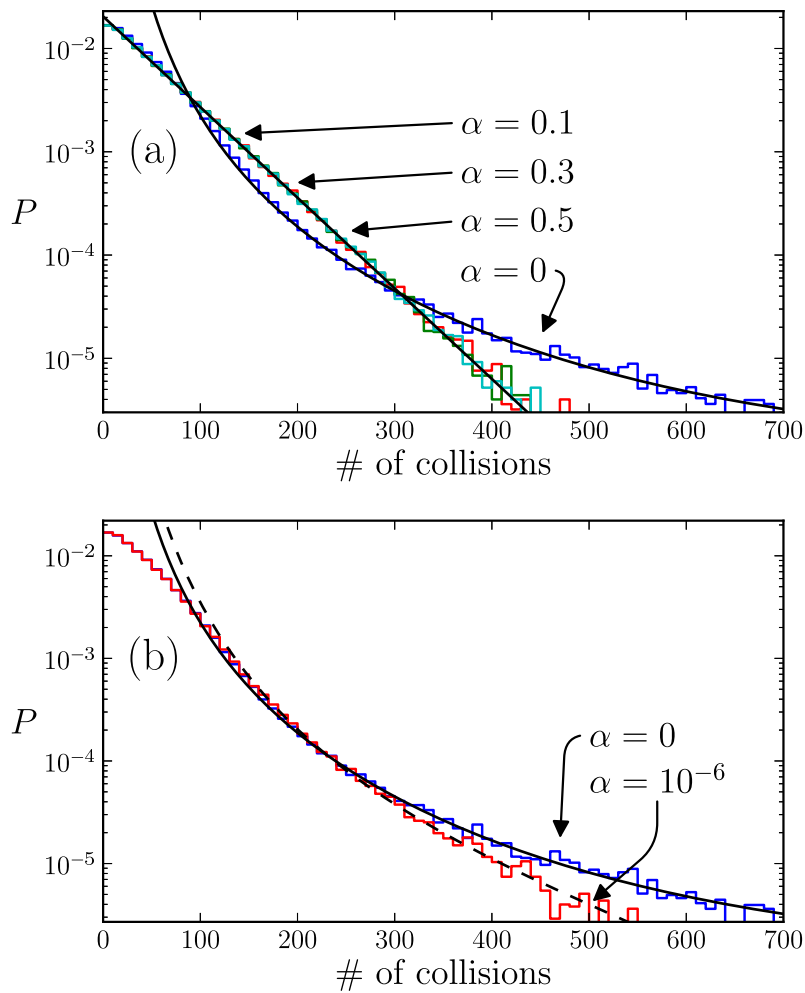}
\caption{(a) Histograms of the escape rates in a two-electron circular billiard.
The noninteracting case shows power-law behavior in the
tail (curved solid line),
whereas at $0.1\leq\alpha\leq 0.5$ the escape rate
is exponential (straight solid line). (b) At very weak interactions
$\alpha=10^{-6}$ we find a mixture of
these tendencies due to quasi-regular trajectories in the system.}
\label{fig5}
\end{figure}

Figure~\ref{fig5}(a) shows the resulting histograms of the escape-rate
calculations. First, the noninteracting situation $(\alpha=0)$ has
a clear power-law tail with $P(n)\propto n^{-\gamma}+{\rm const}$, where
$\gamma\approx 3.46$. In contrast, when $0.1\leq\alpha\leq 0.5$ an excellent
fit to the exponential behavior with $P(n)=49^{n-1}/50^n$
(straight line) can be found. This relation results from the system
geometry: each collision has the escape probability of $1/50$, and
thus for the $n$:th collision to lead
to escape we find $P(n)=(49/50)^{n-1}(1/50)$. This essentially means that
the correlation with two successive bounces is completely lost, and hence
the system can be classified as chaotic.

In Fig.~\ref{fig5}(b) we have a closer look to the weak-interaction limit
with $\alpha=10^{-6}$. A good fit to the computed data is obtained
with a power-law curve having $\gamma\approx 4.08$, i.e., the
escape is slightly faster than in the noninteracting limit. However,
at small time scales the behavior is very similar to the
$\alpha=0$ as the trajectories essentially follow the same
(quasi-)regular patterns. These quasi-stable trajectories also
give arise to power-law tail in the escape-rate distributions
for weak interactions. However, the interaction reduces the lengths
of the quasi-regular parts of the electron trajectories and thus decreases
the survival probability (and escape rates) at longer time scales.

Concluding, our results on the escape probabilities show that
the transition to exponential escape rates is (i) smooth (not abrupt as a function of $\alpha$),
(ii) it occurs first at large times (large number of collisions) in
the histogram, and (iii) it generally appears at relatively small values
for $\alpha$. Our tests indicate that at $\alpha\sim 10^{-3}$ the most
part of the calculated escape-rate histogram is closer to an
exponential behavior than to the power-law one.
We point out that these numerical experiments do not exclude the possibility
of power-law escape rates with intermediate interaction strengths as $n$ tends to infinity.
Also, the large-$\alpha$ regime is excluded in this
analysis due to numerical reasons: at $\alpha>0.5$ we would need
to decrease the size of the holes due to small-scale motion close to the boundary,
and thus the time step should be decreased as well. Hence, for
consistency of the results we have focused here only
on the range $0\leq\alpha\leq 0.5$.

\section{Summary}\label{summary}

To summarize, we have made a thorough look into chaotic dynamics
of circular billiards containing two Coulomb-interacting electrons
with the full range of interaction strengths ($0\leq\alpha\leq 1$).
Close to both weak- and strong-interaction limits the bouncing maps show
traces of quasi-regular behavior, although the dynamics generally
appears as chaotic. In the strong-interaction limit we are able to find
an analytic expression for the bouncing map that
agrees very well with the calculated data at $\alpha\rightarrow 1$.
At smaller $\alpha$ the predictive power of the expression reduces,
although the agreement is reasonable down to $\alpha\sim 0.9$.
To assess the change in dynamics as interaction is increased
we have calculated escape rates as a function of $\alpha$
and found similar exponential behavior
through a wide range of interaction strengths.
Thus, within the examined time scales 
our results suggest universally chaotic behavior
in Coulomb-interacting hard-wall billiards apart from the 
noninteracting and possibly strong-interacting limits.


\ack 
We are grateful to Rainer Klages, Sebastian Schr\"oter, Javier
Madro\~nero, and Paul-Antoine Hervieux for useful discussions.
This work was supported by the Academy of Finland
and the Finnish Cultural Foundation.
We are grateful to CSC -- the Finnish IT Center for
Science -- for computational resources.

\end{document}